\documentclass[journal=inoraj,manuscript=article]{achemso}
\usepackage[version=3]{mhchem}
\usepackage{color}

\title{Short-range Crystalline Order-Tuned Conductivity in Cr$_2$Si$_2$Te$_6$ van der Waals Magnetic Crystals}
\author{Yu Liu,$^{\dag,\sharp,\ast}$ Resta A. Susilo,$^{\P,\flat}$ Yongbin Lee,$^{\diamondsuit}$ A. M. Milinda Abeykoon,$^{\|}$ Xiao Tong,$^{\S}$ Zhixiang Hu,$^{\dag,\ddag}$ Eli Stavitski,$^{\|}$ Klaus Attenkofer,$^{\|,\P}$ Liqin Ke,$^{\diamondsuit}$ Bin Chen,$^{\P}$ and Cedomir Petrovic $^{\dag,\ddag,\ast}$}
\email{yuliu@lanl.gov; petrovic@bnl.gov}
\affiliation[BNL]
{$^{\dag}$Condensed Matter Physics and Materials Science Department, Brookhaven National Laboratory, Upton, New York 11973, USA\\
$^{\P}$Center for High Pressure Science and Technology Advanced Research, Pudong, Shanghai 201203, China\\
$^{\diamondsuit}$Ames Laboratory, U.S. Department of Energy, Ames, Iowa 50011, USA\\
$^{\|}$National Synchrotron Light Source II, Brookhaven National Laboratory, Upton, New York 11973, USA\\
$^{\S}$Center for Functional Nanomaterials, Brookhaven National Laboratory, Upton, New York 11973, USA\\
$^{\ddag}$Materials Science and Chemical Engineering Department, Stony Brook University, Stony Brook, New York 11790, USA
}

\begin{document}

\begin{tocentry}

\centerline{\includegraphics[scale=1]{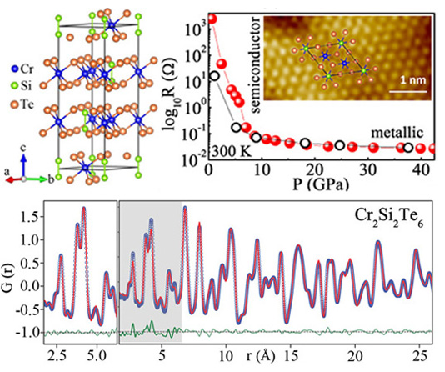}}

\end{tocentry}

\newpage
\begin{abstract}
  Two-dimensional magnetic materials (2DMM) are significant for studies on the nature of 2D long range magnetic order but also for future spintronic devices. Of particular interest are 2DMM where spins can be manipulated by electrical conduction. Whereas Cr$_2$Si$_2$Te$_6$ exhibits magnetic order in few-layer crystals, its large band gap inhibits electronic conduction. Here we show that the defect-induced short-range crystal order in Cr$_2$Si$_2$Te$_6$ on the length scale below 0.6 nm induces substantially reduced band gap and robust semiconducting behavior down to 2 K that turns to metallic above 10 GPa. Our results will be helpful to design conducting state in 2DMM and call for spin-resolved measurement of the electronic structure in exfoliated ultrathin crystals.
\end{abstract}

\section{Introduction}

Intrinsic ferromagnetic (FM) semiconductors are of great interest for fundamental studies of magnetism and for future spintronic technology \cite{Jungwirth}. In particular, due to use in nanoscale devices, significant effort is devoted towards computational predictions and discovery of crystals with layered structure with van der Waals (vdW) bonds \cite{Lebegue}. First-principles calculations, Raman experiments and electronic transport indicated that FM state in Cr$_2$X$_2$Te$_6$ (X = Si, Ge) survives even in mono- and few-layer crystals and indeed this is experimentally observed in bilayer Cr$_2$Ge$_2$Te$_6$ and related CrI$_3$ 2DMMs \cite{Li,Lin,Huang,Gong}. This presents potential opportunity for fundamental studies of 2D magnetism and further magneto-electronic device design. In particular, due to magneto-elastic coupling in Cr$_2$Si$_2$Te$_6$, ferromagnetic spin exchange via Te-to-Cr charge transfer can be manipulated by ultrafast optical pulses that excite coherent phonon oscillations \cite{RonA}. Furthermore, of high interest are conducting electronic states in 2D materials that can be used for spin-charge conversion in spintronic devices \cite{RalphD,BrataasA,WeiH,ACSNANO}.

Cr$_2$Si$_2$Te$_6$ orders ferromagnetically below 32 K and features band gap of 0.4 eV \cite{Carteaux,Casto}. Cr$_2$Ge$_2$Te$_6$ has FM $T_\textrm{c}$ of 61 K and band gap of 0.7 eV \cite{Siberchicot,Zhang}. FM correlations along the $c$-axis vanish above 50 K in Cr$_2$Si$_2$Te$_6$ but in-plane short-range FM correlations persist up to room temperature. In addition, there is strong spin-phonon coupling \cite{Williams}. Whereas spin-phonon interaction is robust on chemical alloying and vacancies, application of high-pressure induces metallic state above 8 GPa and superconductivity \cite{Milosavljevic,Cai}.

On the other hand, different forms of crystalline imperfections have been predicted by the first-principle calculations to tune both magnetic and conducting states in Cr$_2$X$_2$Te$_6$. Surface Ge vacancies lead to metallic state by closing the band gap in Cr$_2$Ge$_2$Te$_6$ multilayer, causing a magnetic anisotropy energy rise \cite{Song}. Strain in Cr$_2$Ge$_2$Te$_6$ monolayer can raise ferromagnetic $T_\textrm{c}$ close to room temperature \cite{Chen}. Of particular interest is theoretical consideration of crystalline imperfection engineering in Cr$_2$Si$_2$Te$_6$ monolayer. Defects are predicted to result in fully spin polarized current arising from bipolar magnetic semiconducting state where reversible spin direction of opposite spin-polarized conduction and valence bands state may be tuned by the gate voltage \cite{Cheng}. This is favored over external magnetic field switching of single spin direction in standard half-metals and semiconductors; moreover bipolar magnetic semiconductors could also be used for separation and detection of entangled states in superconductors, of interest for quantum computing \cite{LiX1,CarusoA,LiX2}.

In this work we successfully synthesized exfoliable crystals of Cr$_2$Si$_2$Te$_6$ with defect-induced short-range crystalline order imperfections on the length scale below 1 nm. This suppresses the 0.4 eV gap in the charge insulating state above magnetic $T_c$ \cite{ZhangJ} and gives rise to semiconducting current that can be tuned to a good metal at high pressure. We observe considerably lower thermal conductivity when compared to previous reports \cite{Casto} arising from possible stacking faults and higher magnetic entropy released below the magnetic transition which implies not only increased phonon scattering but also perturbation of phonon modes and possibly of magnetoelastic coupling. Enhanced conduction suggests that Cr$_2$Si$_2$Te$_6$ exfoliable crystals may be of interest for future spintronic devices, calling for investigation of spin-polarized electronic structure and spin-polarized current in  epitaxially strained  Cr$_2$Si$_2$Te$_6$ nanostructures.

\section{RESULTS AND DISCUSSION}

Cr$_2$Si$_2$Te$_6$ [Fig. 1(a)] was previouly synthesized by Ouvrard \emph{et al.} \cite{Ouvrard}, which is a layered material with Si-Si pairs. Cr atoms make honeycomb lattice in planes separated by vdW bonds with Si in hexagon center. Each Cr is octahedrally coordinated by Te, Si-Si bonds make dimers which form Si$_2$Te$_6$ groups \cite{Ouvrard,Marsh,Carteaux2}. Dimers of Si-Si are a peculiarity when compared with CrI$_3$, whereas different electronegativity of constituent atoms suggests variable character of chemical bonds in planes separated by vdW force. Magnetic interaction in planes are FM, arising by Cr-Te-Cr superexchange through nearly 90$^\circ$ angle and is stronger than antiferromagentic (AFM) Cr-Cr direct exchange. The exchange interactions from the next-nearest- and third-nearest- besides the nearest-next (NN) couplings also contribute to its magnetic ground state \cite{Sivadas}. XRD $2\theta$ scan [Fig. 1(b)] reveal only $(00l)$ peaks, indicating layered structure. XRD powder pattern taken at NSLS II synchrotron is fitted with main phase of $R\overline{3}h$ space group Cr$_2$Si$_2$Te$_6$ with $\sim$ 1.7\% SiTe$_2$ impurity [Fig. 1(c)]. Room temperature unit cell parameters of Cr$_2$Si$_2$Te$_6$ are $a = b = 6.7607(2)$ {\AA} and $c = 20.6650(4)$ {\AA} \cite{Ouvrard}.

Figure 1(d,e) presents the XANES and the corresponding Fourier transform spectra of the EXAFS for the Cr and Te $K$-edge of Cr$_2$Si$_2$Te$_6$, respectively, at the room temperature. The XANES spectra indicate Cr$^{3+}$ and Te$^{2+}$ states in Cr$_2$Si$_2$Te$_6$ \cite{Ofuchi,AFM,YLPRB}. The local structural information is revealed in the EXAFS of Cr$_2$Si$_2$Te$_6$ measured at room temperature. By the EXAFS equation, which in the single-scattering approximation takes the form \cite{Prins}:
\begin{align*}
\chi(k) = \sum_i\frac{N_iS_0^2}{kR_i^2}f_i(k,R_i)e^{-\frac{2R_i}{\lambda}}e^{-2k^2\sigma_i^2}sin[2kR_i+\delta_i(k)].
\end{align*}
Here $N_i$ is the number of surrounding atoms from the central photoabsorbing atom at a distance of $R_i$. $S_0^2$ is the reduction factor for passive electrons, $f_i(k, R_i)$ describes backscattering amplitude, $\delta_i$ and $\lambda$ are phase shift and mean free path of photoelectron, respectively. The $\sigma_i^2$ is Debye-Waller factor. The local chemical bond distances from 1.5 to 4.0 {\AA} (Table I) show that local environment of Cr atoms consists of six Te atoms with two close distances [2.81(9) {\AA} and 2.86(9) {\AA}]; the three Cr atoms are further away [3.91(49) {\AA}]. For the Te site, the NN are one Si [2.35(26) {\AA}] and two Cr with close distances [2.81(27) {\AA} and 2.86(27) {\AA}] whereas two Si [3.68(26) {\AA}] and six Te [3.89(12) {\AA} and 3.94(12) {\AA}] are further away. The obtained results from different element EXAFS analysis agree well with each other. First coordination shell bond distances seen from EXAFS do not indicate difference from previously published bond distances seen by the average crystal structure where Cr NN Te atoms are at 2.81 {\AA} and 2.86 {\AA} away and Te NN atoms are Si at 2.34 {\AA} and Cr at 2.81 {\AA} and 2.86 {\AA} \cite{Carteaux3}.

For further insight into change of interatomic distances beyond first coordination shell we studied scanning tunneling microscope (STM) as well as local and average crystal structure from total scattering. The surface of Te atomic layer is easily exposed by crystal cleaving \cite{Hao}. Figure 1(h) presents the STM topography obtained on the surface of cleaved Cr$_2$Si$_2$Te$_6$ crystal. An ordered honeycomb lattice of bright spots is observed; each spot is a surface Te atom. It should be noted that a large number of darker spots (defects) were also observed [Fig. 1(h) upper panel], which tunes the local electronic structure. Figure 1(i) shows the comparison of the spatially-averaged dI/dV spectrum of bright and dark areas measured at the room temperature. As we can see, there is a small energy gap around $E_\textrm{F}$, which is more narrow in the dark area [Fig. 1(i)]. A similar defects feature was also observed in Cr$_2$Ge$_2$Te$_6$, which is explained by assuming the replacement of Cr by Ge atoms \cite{Hao}. The local energy gap between maximum of valence band and Fermi level is about $\sim$ 0.16(1) eV from the Ultraviolet Photoelectron Spectroscopy (UPS). The bright surface domains/areas, though, appear long range order in atomic structure, but are randomly ``waved" as shown in large scale STM images. The randomly waved/corrugated top atomic layers are causing the interlayer distance to vary randomly, which may cause random variation of the the interlayer electrical coupling in the real space. Thus, in momentum space, there could be changes in band dispersion in the Brillouin zone (BZ) that affect electron correlation. This is similar to twisted or pressed bi- or multi-layer graphene where the interlayer electrical coupling can be dramatically changed by converting linearly dispersive Dirac cones into flat bands at some magic twist angle or pressure, giving rise to correlated electronic liquid that may exhibit skyrmions in ferromagnetic state or unconventional superconductivity \cite{CaoY,Bomerich}. On other hand, the randomly distributed dark areas could act as electron scattering centers, resulting in the Anderson localization, less dispersive and more correlated conducting bands. This implies that the dark features in STM may posses unique crystal structure with different coordinate lattice environment when compared to defects-free surface area.

Theoretical calculations predict Cr(Si) but also Si(Cr) antisite defects that are energetically favorable in monolayer Cr$_2$Si$_2$Te$_6$, which has lower formation energy than other vacancies and/or interstitial defects \cite{Cheng}. The defect engineering has been predicted to give rise to bipolar magnetic semiconducting state with highly spin polarized current; in such state a gate voltage control of reversible spin detection is possible. For insight into nanoscale crystal structure we present the Rietveld and pair distribution function (PDF) result from total scattering. Long-range order (LRO) atomic structure of Cr$_2$Si$_2$Te$_6$ [Fig. 1(c)] features relatively large atomic displacement parameters (ADPs) (see Supplementary Materials) and the noticeable diffuse scattering component in background subtracted data suggested short-range ordered (SRO) structural modulations. The PDF analysis was further performed to confirm this observation. Not surprisingly, LRO model can explain crystal structure on the (6.2 - 25) {\AA} length-scale but shows a rather poor agreement on (1.5 - 6.2) {\AA} of PDF refinement [Fig. 1(k)]. Further refinement of the LRO model on the PDF length scale (1.5 - 6.2) {\AA} resulted in noticeably different bond lengths and angles. SRO at distances below 6.2 {\AA} is well fitted with the same trigonal symmetry but with somewhat different unit cell dimensions $a = b = 6.76151(1)$ {\AA} and $c = 20.6138(1)$ {\AA} that introduce changes in local environment of Cr atoms. The detailed refinement can be found in Supplementary Materials that is considered as a distorted structure in the DFT calculations (see below).

The $M(T)$ of our Cr$_2$Si$_2$Te$_6$ crystal is nearly isotropic above 40 K for both directions but a large magnetic anisotropy sets in below 40 K. For $\mathbf{H\parallel ab}$, a clear two-step transition occurs, not present in $M(T)$ measured in higher fields \cite{YuLiu}, and the splitting of ZFC and FC was observed below 26 K. To probe anisotropy of the two-stage transition \cite{XieQ}, we performed ac magnetization measurements. An obvious kink (satellite transition) at $T^*$ $\sim$ 26 K was further confirmed in real part of ac susceptibility $m'(T)$ in addition to the upturn around $T_\textrm{c}$ $\sim$ 35 K [Fig. 2(a) inset]. For $\mathbf{H\parallel c}$, the ZFC - FC splitting below $T_\textrm{c}$ is due to the anisotropic FM domain effect. The clear peak at $T_\textrm{c}$ $\sim$ 35 K was also verified in $m'(T)$ [inset in Fig. 2(b)]. A clear $\lambda$-type peak in heat capacity [Fig. 2(c)] corresponds to the ferromagnetic transition observed at 37 K in zero field. The $C_\textrm{p}(T)$ below 12 K can be described by $C_\textrm{p}(T) = \gamma T+ \beta T^3 + \delta T^{3/2}$, where first two terms describe electronic and phonon contributions and $\delta T^{3/2}$ is the spin-wave contribution \cite{Gopal}. The $\gamma$ is negligible, $\beta$ = 1.18(3) mJ mol$^{-1}$ K$^{-4}$ and $\delta$ = 36(1) mJ mol$^{-1}$ K$^{-5/2}$. The Debye temperature from $\Theta_\textrm{D} = (12\pi^4NR/5\beta)^{1/3}$ = 254(2) K with $N = 10$. We obtain magnetic entropy $S(T) = \int_0^T C_p(T,H)/TdT$ = 3.1 J mol$^{-1}$ K$^{-1}$ in (25-45) K range, higher when compared to previously observed but still far below expected $R$ln(2$S$+1) = 11.53 Jmol$^{-1}$K$^{-1}$ expected total magnetic entropy \cite{Casto}. The magnetic entropy change $\Delta S_\textrm{M}(T,H) = S_\textrm{M}(T,H)-S_\textrm{M}(T,0)$, assuming adiabatic process of field change and electronic and phonon contributions are field independent. The $\Delta T_{\textrm{ad}}$, adiabatic temperature change induced by magnetic field change is $\Delta T_{\textrm{ad}}(T,H) = T(S,H)-T(S,0)$. Here $T(S,H)$ and $T(S,0)$ are the temperatures at constant total entropy $S$ in magnetic field and in the absence of magnetic field, respectively. $-\Delta S_\textrm{M}$ and $\Delta T_{\textrm{ad}}$ [Fig. 2(d)] show maximum around $T_c$ and rise with magnetic field increase. $-\Delta S_\textrm{M}$ and $\Delta T_{\textrm{ad}}$ show maximum values of 6.77 J kg$^{-1}$ K$^{-1}$ and 3.0 K in 9 T, somewhat higher when compared to magnetic entropy change estimated from magnetization curves up to 5T \cite{YuLiu}, and also larger than those of CrX$_3$ (X = Cl, Br, I) and Cr$_2$Ge$_2$Te$_6$ \cite{ACSNANO}.

Thermal conductivity $\kappa(T)$ [Fig. 2(e)] is smaller below room temperature from previously observed \cite{Casto}. The $\kappa(T)$ is contributed by the electronic term $\kappa_\textrm{e}$ and the lattice part $\kappa_{\textrm{L}}$, i. e., $\kappa = \kappa_\textrm{e} + \kappa_{\textrm{L}}$. From Wiedemann-Franz law we can calculate the $\kappa_\textrm{e} = L_0T/\rho$ from the measured resistivity $\rho$ with $L_0$ = 2.45 $\times$ 10$^{-8}$ W $\Omega$ K$^{-2}$. As depicted in the inset in Fig. 2(e), $\kappa_\textrm{e}$ is about ten times smaller than $\kappa_{\textrm{L}}$. An anomaly was observed at $T_\textrm{c}$ (red dashed line) for both $\kappa_\textrm{e}$ and $\kappa_{\textrm{L}}$, confirming the strong spin-phonon coupling \cite{Casto,Milosavljevic}. Magnetic field-dependent $\kappa(T)$ in Cr$_2$Si$_2$Te$_6$ exhibits visible deviation from the zero-field $\kappa$($T$) due to magnetic correlations above $T_\textrm{c}$, in agreement with entropy \cite{Casto} and neutron scattering results \cite{Williams}. $\kappa(T)$ in [Fig. 2(e)] increases in magnetic field, suggesting possible phonon scattering off RuCl$_3$-like Kitaev-Heisenberg excitations as discussed in Cr$_2$X$_2$Te$_6$ ($X$ = Si, Ge) \cite{Hentrich,XuC1,XuC2}. The $S(T)$ [Fig. 2(f)] is positive with $S$(300 K) =180 $\mu$V K$^{-1}$. The $S(T)$ is almost linear in high temperature. Assuming energy-independent relaxation time in the free electron model $S = \pi^2k_\textrm{B}^2T/3eE_\textrm{F}$. Hence, fit result $dS/dT$ = 0.492(6) and 0.526(3) $\mu$V K$^{-2}$ derives the Fermi energy $E_\textrm{F}$ = 49.7(6) and 46.4(3) meV in 0 and 9 T, respectively. The $S(T)$ shows a kink at $T_\textrm{c}$ $\sim$ 34 K in zero magnetic field, in agreement with the $\kappa(T)$ measurement, which is suppressed in 9 T. The low temperature $S(T)$ is well fitted by $S(T) = (k_\textrm{B}/e)(A+E_\textrm{S}/k_\textrm{B}T) + BT^{1/2}$ [inset in Fig. 2(f)] \cite{Austin,Zvyagin}, where $E_\textrm{S}$ $\sim$ 30(1) $\mu$eV is the activation energy whereas $A$ is constant, the second term is attributed to the variable-range hopping (VRH) conduction.

Resistivity $\rho(T)$ of Cr$_2$Si$_2$Te$_6$ at ambient pressure is presented in Fig. 3(a). It increases on cooling from the room temperature, showing a typical semiconducting behaviour. Below 50 K, $\rho(T)$ increases faster with a weak shoulder at $T_\textrm{c}$. The temperature-derivative curve shows an anomaly at 37 K, consistent with the $T_\textrm{c}$ observed from the magnetization and specific heat data. Therefore the weak shoulder in the $\rho(T)$ data is associated with the FM transition. Above 40 K, resistivity is explained by the thermally activated model in the presence of disorder-induced localization of the electronic wavefunction: $\rho(T)$ = $\rho$$_0$ + $A$exp($E_\rho$/$k_\textrm{B}$$T$) - $B$$T^{0.5}$, where $\rho$$_0$ is residual resistivity, prefactors $A$ and $B$ denote relative contribution of activated and localization scattering mechanism whereas $E_\rho$ is the band gap \cite{Lee,SinghS}. The fit yields $E_\rho$ $\sim$ 18.0(1) meV activation energy, $A$ = 10(2)$\times$10$^{-5}$ and $B$ = 1.77(3)$\times$10$^{-3}$, which shows that electronic transport is dominated by disorder-induced localization. The $\rho(T)$ below magnetic transition can be fitted with the same equation used for paramagnetic state but where band conduction yields to adiabatic small polaron hopping, i.e. temperature activated term $A$exp($E_\rho$/$k_\textrm{B}$$T$) is replaced by $CT$exp($E$$_\rho$/$k_\textrm{B}$$T$) \cite{Austin}. The derived $E_\rho$ $\sim$ 0.30(2) meV, $C$ = 2.3(2)$\times$10$^{-3}$ and $D$ = 43(1)$\times$10$^{-3}$. The $E_\textrm{S}$ $<$ $E_\rho$ is consistent with polaron contribution to transport mechanism, since $E_\textrm{S}$ is associated with carrier hopping, whereas energy to both create and activate the hopping of carriers are in $E_\rho$ \cite{Austin}. It should be noted, though, that disorder-induced localization is dominant in paramagnetic state as well below magnetic order, which shows high correlation between electronic transport and inhomogeneous nanoscale short range crystallographic distortions.

The electronic transport in Cr$_2$Si$_2$Te$_6$ can be further tuned by pressure. The pressure-dependent resistance at 300 K [Fig. 3(a) inset] significantly drops from $10^3$ $\Omega$ below 1 GPa to values less than $10^{-1}$ $\Omega$ around 10 GPa and above. The giant drop in resistance signals pressure-induced semiconductor-metal transition as observed in the isostructural Cr$_2$Ge$_2$Te$_6$ \cite{Yu}. This critical value of pressure ($\sim$ 10 GPa) agrees well with the theoretical prediction of pressure-induced semiconductor-metal transition ($\sim$ 10.4 GPa) in distorted Cr$_2$Si$_2$Te$_6$ (see below). In order to confirm this semicondutor-metal transition, we measured the temperature-dependent data at different pressures [Fig. 3(b)]. At 5.7 GPa, a hump in the resistivity emerges, proposed to be related to the insulator-metal transition (IMT) \cite{Cai}. On increasing pressure, the hump shifts towards high temperature and simultaneously weakens in high pressure. The resistivity behavior above 9 GPa suggests that Cr$_2$Si$_2$Te$_6$ approaches a metallic state. Indeed, the resistivity at 12.5 GPa falls below the Mott-Ioffe-Regel (MIR) limit~\cite{mott, kurosaki}, $\rho_{\textrm{MIR}} = \hbar c/e^2 ~\sim~9\times10^{-4}~\Omega$ cm, where $c$ is the lattice constant ($\sim20.5$~\AA). Above 20 GPa, the resistance decreases with decreasing temperature which indicates the metallization of compressed Cr$_2$Si$_2$Te$_6$. Note that no obvious resistivity drop, which signals superconducting transition, was observed from the resistivity data measured above 9 GPa. This is in contrast with the pressure-induced superconductivity in Cr$_2$Si$_2$Te$_6$ above $\sim$ 8 GPa reported previously, implying that short-range structural details [Fig. 1(k)] are detrimental for the high-pressure superconducting state \cite{Cai}.

Structural properties of Cr$_2$Si$_2$Te$_6$ at high pressures have been investigated recently by Cai \textit{et al.} \cite{Cai} and Xu \textit{et al.} \cite{xu}. However, both reports arrived at different conclusion regarding whether Cr$_2$Si$_2$Te$_6$ transforms into another high pressure phase on compression. Here, we used Raman spectroscopy to probe the vibrational properties of Cr$_2$Si$_2$Te$_6$ under pressure up to 20 GPa connected with the putative structural transition Fig. 3(c). It is known that many Te-containing materials experience easy surface oxidation, which is also problematic in the case of Cr$_2$Si$_2$Te$_6$ since Te-oxide (TeO$_x$) shows very strong Raman signals. In order to minimize surface oxidation, we used a freshly cleaved crystal in glovebox and attempted to load the crystal into the diamond anvil cell in less than 5 minutes. As shown in Fig. 3(d), a contribution to the Raman spectra of Cr$_2$Si$_2$Te$_6$ due to the TeO$_x$ impurities can be excluded. Group theory analysis revealed that the Raman tensor is represented by five $A_g$ and five doubly degenerate $E_g$ modes, \textit{i.e.} $\Gamma = 5A_g + 5E_g$ \cite{milosav}. Raman spectrum of Cr$_2$Si$_2$Te$_6$ at smallest pressure of 0.3 GPa features two peaks at around 117~cm$^{-1}$ and 147~cm$^{-1}$ which are assigned to $E_g^3$ and $A_g^3$ modes, respectively [Fig. 3(d)]. Both the $E_g^3$ and $A_g^3$ modes shift to high frequencies with increasing pressure without apparent anomalies [Fig. 3(e,f)]. No peaks splitting or the appearance of new Raman mode was observed, implying no crystal structure changes in Cr$_2$Si$_2$Te$_6$ up to 18 GPa, consistent with the high pressure XRD data by Xu \textit{et al.} \cite{xu}. The intensities of both modes weaken above 12 GPa and completely disappear at 20 GPa probably due to metallization of Cr$_2$Si$_2$Te$_6$ as revealed from the resistance data in Fig. 3(b). It is also possible that the Raman modes vanishing is due to the loss of long-range order due to the formation of amorphous phase reported to occur above 24 GPa \cite{xu}. Given that Raman spectroscopy probes local atomic vibrations, thus it is also sensitive to the short range crystalline structure, we also carried out another run on another batch of sample prepared with the same method. The evolution of $E_g^3$ and $A_g^3$ modes [Fig. 3(e,f)] is consistent between different samples which indicates that such short range crystalline order is reproducible in our grown samples.

To comprehend the implications of structural distortion, we further performed the band structure calculation as well as calculations of conductivity in undistorted and distorted (SRO) structures. The band structures [Fig. 4(a)] show flat bands with similar energies between $T$ and $\Gamma$, and the maximum of valence band (VBM) is in between $T$ and $\Gamma$ points. In comparison, the valence bands of the undistorted structure have higher energy than the bands of the distorted structure near the Fermi energy ($E_\text{F}$). For the conduction band, while the undistorted structure shows a minimum in conduction band (CBM) between H$_{2}$ and $T$ Brillouin zone points, the distorted one shows three CBMs between H$_{2}$ and $T$, between $L$ and H$_{0}$, and between $\Gamma$ and S$_{0}$ points. The CBMs between H$_{2}$ and $T$ have similar energy in both structures, and these bands are majority-spin dominant bands. The other two CBMs of the distorted structure are minority-spin dominant bands, and corresponding bands of the undistorted structure have higher energies. The undistorted structure has a bandgap of 0.45 eV without spin-orbit coupling (SOC); including SOC decreases the gap size to 0.27 eV. Similar SOC effects on bandgap have been found in other magnetic 2D vdW materials \cite{lee2020prb}. The distorted structure has slightly different bandgaps that are 0.47 eV and 0.25 eV without and with SOC, respectively. These bandgaps are all indirect.

To understand the pressure-induced IMT in Cr$_2$Si$_2$Te$_6$, we investigated how electronic structures and transport properties evolve with the decreasing of lattice constants using DFT. For simplicity, we keep the $c/a$ constant (adopted from the experimental value) while changing the volume. Compared to the experimental values at ambient pressure, DFT slightly overestimates the lattice parameters by $1.8\%$ and $1.2\%$ in the undistorted and distorted structures, respectively. To qualitatively relate the volume change to pressure and compare it with experiments, we also fit the $E$(V) curve, where $E$ is the total energy and $V$ is the volume, by Birch-Murnaghan equation of state (EOS) \cite{Murnaghan244}. Figure 4(b) shows the calculated bandgap as functions of volume in undistorted and distorted Cr$_2$Si$_2$Te$_6$. With decreasing lattice parameters, the bandgap decreases in the distorted structure, while the gap in the undistorted structure slightly increases and then decreases. The semiconductor-metal transition occurs at $\Delta V=-10\%$ and $-15\%$ for the distorted and undistorted structures, respectively. This volume change corresponds to about the pressure value of 10.4 (18.6) GPa for the distorted (undistorted) structure. The calculated transition pressure of 10.4 GPa for the distorted structure agrees well with the experimental observed value. This shows that local distortion can affect the semiconductor-metal transition pressure.

Figure 4(c) shows the spin-projected density of states (DOS) calculated with various volumes in undistorted structure. The minority spin band gap is larger than the majority spin band gap at ambient pressure. However, the majority- and minority-spin bands respond to pressure differently. The majority-spin bandgap slightly increases with increasing pressure while the minority-spin gap decreases. This results in the bandgap increase with pressure in a small pressure region until the minority-spin bandgap becomes smaller than the majority-spin bandgap (see Fig. S2). The actual bandgap is determined by the smaller bandgap in two spin channels. Pressure promotes the hybridization between Cr and Te atom by reducing inter-atomic distance and the delocalization of Cr-$d$ states \cite{Menichetti2019_2dmat}. Cr-$d$ states are the dominant states filling the gap. In comparison to non-SOC calculations, SOC reduces the size of the bandgap, resulting in a smaller IMT transition pressure value (see Fig. S2). Figure 4(d) shows the spin-projected DOS in the distorted structure. Compared to the undistorted structure, the distorted structure has a slightly smaller gap in the majority spin, and a much smaller gap in the minority spin channel. As a result, the distorted structure has a lower IMT transition pressure than the undistorted one. Since electronic conductivities adopt changes in electronic band structures, they display a similar character of IMT transition with increasing pressure. For example, at 20 K, the conductivity in distorted structure with $\Delta V=-12\%$ and the undistorted structure with $\Delta V=-16\%$ increase by  $\sim 180$ and $\sim 108$ times, respectively, compared to their ambient structures (see Fig. S4).

The present plain DFT description of the pressure effects on bandstructure and the calculated IMT transition pressure compare well with experimental results. Future theoretical investigation can explore the pressure-dependent structures and bandgaps with more comprehensive approaches. For example, more accurate descriptions of the pressure-dependent structure of vdW materials can be obtained by employing proper vdW density functionals \cite{mcguire2015cm}. However, the best choice of vdW functional often depends on the material investigated, and a more comprehensive structure study can be helpful. Moreover, it will also be interesting to use more sophisticated \textit{ab-initio} methods beyond DFT \cite{lee2020prb} for a better description of the bandgap.

\section{Conclusions}

Defects, dislocations and other crystalline imperfections in 2D materials beyond graphene often bring interesting physics and useful functionality \cite{ZouX}. In this work we have experimentally observed defects-induced short-range crystallographic order on the length scale 0.6 nm in van der Waals magnet Cr$_2$Si$_2$Te$_6$ where conductivity is dominated by polaron hopping, and also discussed by the DFT calculations. Moreover, metallic state emerges at high pressure without change of the crystal lattice symmetry. It is of high interest to further nanofabricate ultrathin crystals of Cr$_2$Si$_2$Te$_6$ with short range order for transport or to tune its properties by epitaxial strain in thin films which could be used for spin-charge conversion in nanoscale devices.

\section{Material and Methods}

Cr$_2$Si$_2$Te$_6$ crystals with defects were made using flux method \cite{YuLiu}. Cr$_2$Si$_2$Te$_6$ stoichiometry of raw materials was sealed in fused silica tube in about 2 Torr Argon gas pressure. The tube was heated to 1100 $^\circ$C in 20 h. After keeping the tube at 1100 $^\circ$C for 3 h, it was cooled to 680 $^\circ$C in 1 $^\circ$C/h. Crystals were decanted in centrifuge from flux and immediately quenched in ice water. The transport and magnetic properties as well as high pressure Raman data are reasonably repeatable in samples from two different batches. The average chemical composition was investigated by energy-dispersive x-ray spectroscopy (EDS) on several points on the crystal surface using JEOL LSM-6500 scanning electron microscope (SEM). The result was consistent with 1 : 1 : 3 stoichiometry. Wide angle X-ray scattering and PDF experiments used pulverized sample in a capillary in transmission mode and 74.5 keV ($\lambda$ = 0.1665 ${\AA}$) of 28-ID-1 (PDF) beamline of the National Synchrotron Light Source II (NSLS II) at Brookhaven National Laboratory (BNL). The data were integrated using pyFAI software, and reduced to obtain experimental PDFs (Q$_{\textrm{max}}$ = 25 {\AA}$^{-1}$) using the xPDFsuite whereas Rietveld and PDF analysis were performed by GSAS-II and PDFgui software \cite{Kieffer,Yang,Toby,Farrow}.

Scanning tunneling microscope (STM) experiments were performed at room temperature in Scienta Omicron VT STM XA 650 with Matrix SPM Control System in $\sim$ 2 $\times$ 10$^{-10}$ torr UHV chamber on a fresh surface obtained by cleaving. The topography was obtained with positive bias in the constant current mode. STM images were analyzed using SPIP software. In scanning tunneling spectroscopy (STS) experiments, the current (I)-voltage (V) curves are recorded at selected spots while the feedback loop is turned off. Each I-V curve is the average of about 40 individual I-V curves, and then the dI/dV curve was generated. For Ultraviolet Photoelectron Spectroscopy (UPS) measurements a $<$ 2 $\times$ 10$^{-9}$ Torr UHV system was used and hemispherical electron energy analyzer (SPECS, PHOIBOS 100) and ultraviolet source (SPECS, 10/35). He(I) at 21.2 eV radiation was used. The spectra were calibrated by the Fermi edge of Au(111) crystal.

X-ray absorption spectroscopy (XAS) was done in fluorescence mode at 8-ID beamline of the NSLS II at BNL. XANES (X-ray absorption near edge structure spectra) and EXAFS (extended X-ray absorption fine structure spectra) analysis was done by Athena software. EXAFS signal was $k^2$-weighed and Fourier-transformed from 2 to 10 {\AA}$^{-1}$ range to analyze data in $R$ space.

Magnetic, thermal and transport measurements were carried out in MPMS-XL5 and PPMS-9. Electrical transport of Cr$_2$Si$_2$Te$_6$ at high pressure were measured using a Be-Cu diamond anvil cell (DAC) of 300 $\mu$m cullet size. A sample chamber and insulating layer between gasket and electrodes was a mix of cubic boron nitride (cBN) and epoxy. Measurements were done on 60 $\mu$m $\times$ 15 $\mu$m crystal using NaCl powder pressure medium which gives a quasi-hydrostatic condition. For electrical resistance in DAC, Pt foil electrodes and van der Pauw four probe method was used. The pressure was calibrated by Ruby \cite{mao}. Renishaw in Via Raman system and 532 nm laser wavelength were used to obtain Raman spectra at 300 K in a 300 $\mu$m symmetric DAC. Small single crystal with few Ruby balls with silicone oil as a pressure medium were put in a 120 $\mu$m hole of stainless steel gasket.

Band structures were calculated using a full-potential linear augmented plane wave (FP-LAPW) in \textsc{Wien2K} \cite{wien2k} with GGA approximation \cite{perdew1996prl}. We used $R_\text{MT}K_\text{max}$ = 8.0 with muffin-tin (MT) radii $R_\text{MT}$ =2.3, 1.8, and 2.1 a.u. for Te, Si, and Cr, respectively, to generate self-consistent potential and charge. We used 371 $k$-points in the irreducible Brillouin zone, iterated until total energy difference is less than 0.01 mRy. SOC is induced in the second-variational method. Transport properties are calculated within the Boltzmann theory through a postprocessing procedure, as implemented in \textsc{BoltzTraP} \cite{madsen2006cpc}, using the eigenenergies of a self-consistent \textsc{wien2k} calculation. The approximation of isotropic relaxation time ($\tau$) is employed.

\section{Supplementary materials}
The Support Information is available free of charge online.

Further details of crystal structure, STM analysis, and DFT calculations.

\section{Author information}

\subsection{Corresponding Authors}
Email:yuliu@lanl.gov\\
Email:petrovic@bnl.gov\\

\subsection{Present Addresses}

$^\sharp$ Los Alamos National Laboratory, Los Alamos, NM 87545, USA\\
$^\flat$ Department of Physics, Pohang University of Science and Technology, Pohang 37673, Korea\\
$^\P$ ALBA Synchrotron Light Source, Cerdanyola del Valles, E-08290 Barcelona, Spain\\

\subsection{ORCIDs}

Yu Liu: 0000-0001-8886-2876\\
Resta A. Susilo: 0000-0003-0799-7416\\
A. M. Milinda Abeykoon: 0000-0001-6965-3753\\
Xiao Tong: 0000-0002-5567-9677\\
Klaus Attenkofer: 0000-0003-1588-3611\\
Liqin Ke: 0000-0001-5130-9006\\
Cedomir Petrovic: 0000-0001-6063-1881\\

\subsection{Author Contribution}
Y.L. (Yu Liu), R.A.S. and Y.L. (Yongbin Lee) contributed equally to this work. C.P. and B.C. conceive the research and designed the experiment. Y.L. (Yu Liu), Z.H., E.S., K.A. and C.P. performed synthesis, transport, thermal, X-ray absorbtion, magnetic measurements and data analysis. X.T. contributed STM studies. M.A. contributed synchrotron powder X-ray Rietveld and PDF measurements and analysis. R.S. and B.C. contributed high pressure Raman, X-ray and transport measurements and analysis. Y.L. (Yongbin Lee) and L.K. contributed the first principle calculations and analysis. Y.L. (Yu Liu) and C.P. wrote the paper with contribution from all authors. All authors discussed the results and commented on the manuscript.

\subsection{Notes}
The authors declare no competing financial interest.

\begin{acknowledgement}
Work at Brookhaven National Laboratory (BNL) is supported by the Office of Basic Energy Sciences, Materials Sciences and Engineering Division, U.S. Department of Energy (DOE) under Contract No. DE-SC0012704. This research used the 28-ID-1 and 8-ID beamlines of the NSLS II, a U.S. DOE Office of Science User Facility operated for the DOE Office of Science by BNL under Contract No. DE-SC0012704. This research used resources of the Center for Functional Nanomaterials (CFN), which is a U.S. DOE Office of Science Facility, at BNL under Contract No. DE-SC0012704. First principle calculations were supported by the U.S. Department of Energy, Office of Science, Office of Basic Energy Sciences, Materials Sciences and Engineering Division. Ames Laboratory is operated for the U.S. Department of Energy by Iowa State University under Contract No. DE-AC02-07CH11358.
\end{acknowledgement}

\pagebreak

\begin{table*}\centering
\caption{\label{tab}Structural parameters obtained from the synchrotron powder XRD and local bond distances extracted from the EXAFS spectra fits with fixed CN for Cr$_2$Si$_2$Te$_6$. CN is the coordination number based on crystallographic value, R is the interatomic distance, and $\sigma^2$ is the Debye Waller factor.}
\begin{tabular}{llllll}
  \hline
  \hline
  \multicolumn{1}{c}{Chemical:} &\multicolumn{2}{c}{Cr$_{2}$Si$_2$Te$_6$} &\multicolumn{2}{c}{Space group:} &\multicolumn{1}{c}{$R\bar{3}h$}\\
  \multicolumn{1}{c}{$a$ ({\AA})} &\multicolumn{2}{c}{6.7607(2)} &\multicolumn{2}{c}{$\alpha$ ($^\circ$)} &\multicolumn{1}{c}{90}\\
  \multicolumn{1}{c}{$b$ ({\AA})} &\multicolumn{2}{c}{6.7607(2)} &\multicolumn{2}{c}{$\beta$ ($^\circ$)} &\multicolumn{1}{c}{90}\\
  \multicolumn{1}{c}{$c$ ({\AA})} &\multicolumn{2}{c}{20.6650(4)} &\multicolumn{2}{c}{$\gamma$ ($^\circ$)} &\multicolumn{1}{c}{120}\\
  \multicolumn{1}{c}{$V$ ({\AA}$^3$)} &\multicolumn{2}{c}{817.98(3)} &\multicolumn{2}{c}{Weight fraction} &\multicolumn{1}{c}{98.3\%}\\
  \hline
   Site & $x$ & $y$ & $z$ & Occ. & U$_{iso}$ ({\AA}$^2$) \\
  \hline
   Te & -0.0028(4) & 0.3519(3) & 0.0832(1) & 1 & 0.0147(2) \\
   Si & 0 & 0 & 0.0602(6) & 1 & 0.027(3) \\
   Cr & 0 & 0 & 0.3385(3) & 1 & 0.012(2)\\
  \hline
  Center & Distance & CN & R ({\AA}) & $\Delta$R ({\AA}) & $\sigma^2$ ({\AA}$^2$)\\
  \hline
  Cr & Cr-Te.1 & 3 & 2.81 & 0.09 & 0.005 \\
     & Cr-Te.2 & 3 & 2.86 & 0.09 & 0.005 \\
     & Cr-Cr.1 & 3 & 3.91 & 0.49 & 0.036 \\
  \hline
  Te & Te-Si.1 & 1 & 2.35 & 0.26 & 0.011 \\
     & Te-Cr.1 & 1 & 2.81 & 0.27 & 0.001 \\
     & Te-Cr.2 & 1 & 2.86 & 0.27 & 0.001 \\
     & Te-Si.2 & 1 & 3.68 & 0.26 & 0.011 \\
     & Te-Te.1 & 4 & 3.89 & 0.12 & 0.008 \\
     & Te-Te.3 & 2 & 3.94 & 0.12 & 0.008 \\
  \hline
  \hline
\end{tabular}
\end{table*}

\begin{figure*}
\centerline{\includegraphics[scale=1]{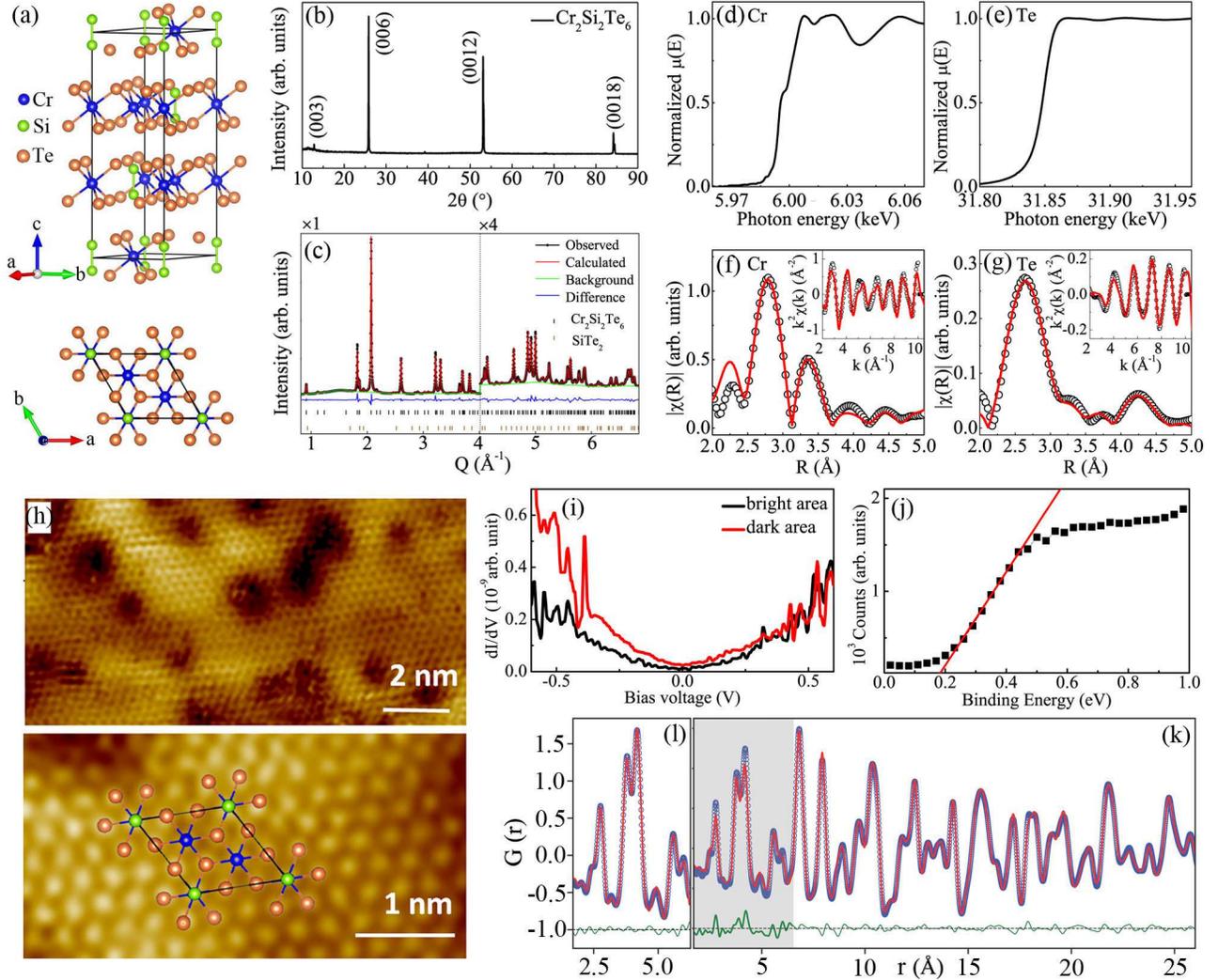}}
\caption{(Color online) (a) Crystal structure of Cr$_2$Si$_2$Te$_6$ (space group: $R\bar{3}h$) shown from the side and top views. (b) Single crystal X-ray diffraction (XRD) and (c) refinement of synchrotron powder XRD data of Cr$_2$Si$_2$Te$_6$ at room temperature. Normalized (d) Cr and (e) Te $K$-edge X-ray absorption near edge structure (XANES) spectra. Fourier transform magnitudes of the extended X-ray absorption fine structure (EXAFS) oscillations (symbols) for (f) Cr and (g) Te $K$-edge with the phase shifts correction. The model fits are shown as solid lines. Insets show the corresponding filtered EXAFS (symbols) with $k$-space model fits (solid lines). (h) Scanning tunneling microscope (STM) images. Upper: sample bias of 1 V and tunneling current of 0.8 nA; lower: 1.3 V and 0.8 nA, respectively. Individual atoms are represented by green (Cr), blue (Si) and brown (Te) colors. (i) Comparison of tunneling spectroscopy at the bright (defect-free) and dark (defect) areas. (j) Ultraviolet Photoelectron Spectroscopy of Cr$_2$Si$_2$Te$_6$ at room temperature. (k) LRO model obtained from the Rietveld refinement showed a good agreement on the PDF length scale (6.2 - 25) {\AA}. However, it gave a poor agreement on the PDF length scale (1.5 - 6.2) {\AA}. (l) Structural model fit on PDF length scale (1.5 - 6.2) {\AA} suggested modulated bond lengths and angles. Blue circles, red, and green solid lines represent data, structural model fits, and the fit residues respectively.}
\end{figure*}

\begin{figure*}
\centerline{\includegraphics[scale=1]{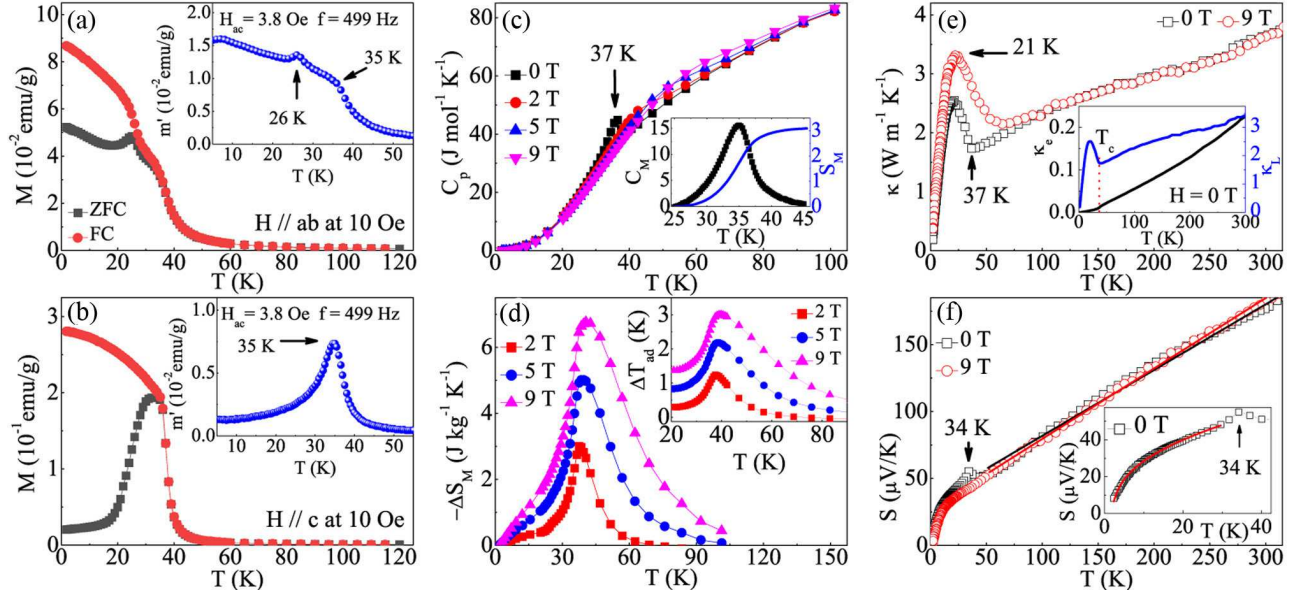}}
\caption{(Color online) Temperature dependence of zero field cooling (ZFC) and field cooling (FC) dc magnetization $M(T)$ measured at $H$ = 10 Oe for Cr$_2$Si$_2$Te$_6$ single crystal with (a) $\mathbf{H\parallel ab}$ and (b) $\mathbf{H\parallel c}$, respectively. Insets show the real part $m^\prime(T)$ of ac susceptibility measured with oscillated an ac field of 3.8 Oe and frequency of 499 Hz. Temperature dependence of (c) specific heat $C_p(T)$ and (d) calculated magnetic entropy change $-\Delta S_M(T)$ in various out-of-plane fields of Cr$_2$Si$_2$Te$_6$ single crystal. Inset in (c) shows magnetic specific heat and magnetic entropy released around the transition in J mol$^{-1}$ K$^{-1}$. Inset in (d) shows the estimated adiabatic temperature $\Delta T_{ad}(T)$. Temperature dependence of (e) thermal conductivity $\kappa(T)$ and (f) thermopower $S(T)$ of Cr$_2$Si$_2$Te$_6$ single crystal in zero field and 9 T. Inset in (e) shows the calculated electronic thermal conductivity $\kappa_e(T)$ and lattice part $\kappa_L(T)$ in W m$^{-1}$ K$^{-1}$. Inset in (f) shows low-temperature $S(T)$ fit (see text).}
\end{figure*}

\begin{figure}
\centerline{\includegraphics[scale=1]{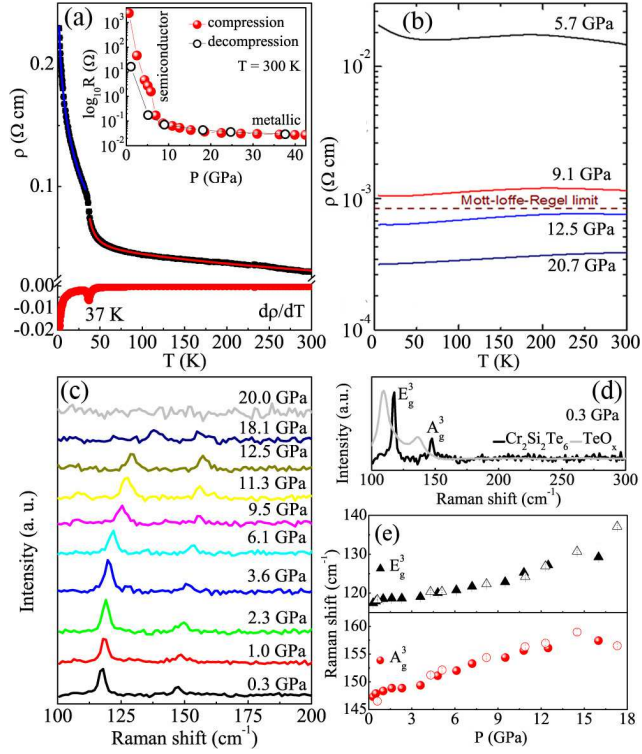}}
\caption{(Color online) (a) Temperature dependence of in-plane electrical resistivity $\rho(T)$ (top) and $d\rho/dT$ (bottom) in zero field of Cr$_2$Si$_2$Te$_6$ single crystal. Inset exhibits the room temperature resistance of a bulk Cr$_2$Si$_2$Te$_6$ sample during compression and decompression. (b) Temperature dependence of $\rho(T)$ at indicated pressures. (c) Raman spectra of Cr$_2$Si$_2$Te$_6$ single crystal at various pressures. (d) Raman spectra of Cr$_2$Si$_2$Te$_6$ compared with that of TeO$_x$ at 0.3 GPa. Pressure-dependent Raman shift of (e) $E^3_g$ and (f) $A^3_g$ modes of two different batches of sample (solid and open symbols represent Samples 1 and 2, respectively).}
\end{figure}

\begin{figure*}
\centerline{\includegraphics[scale=1]{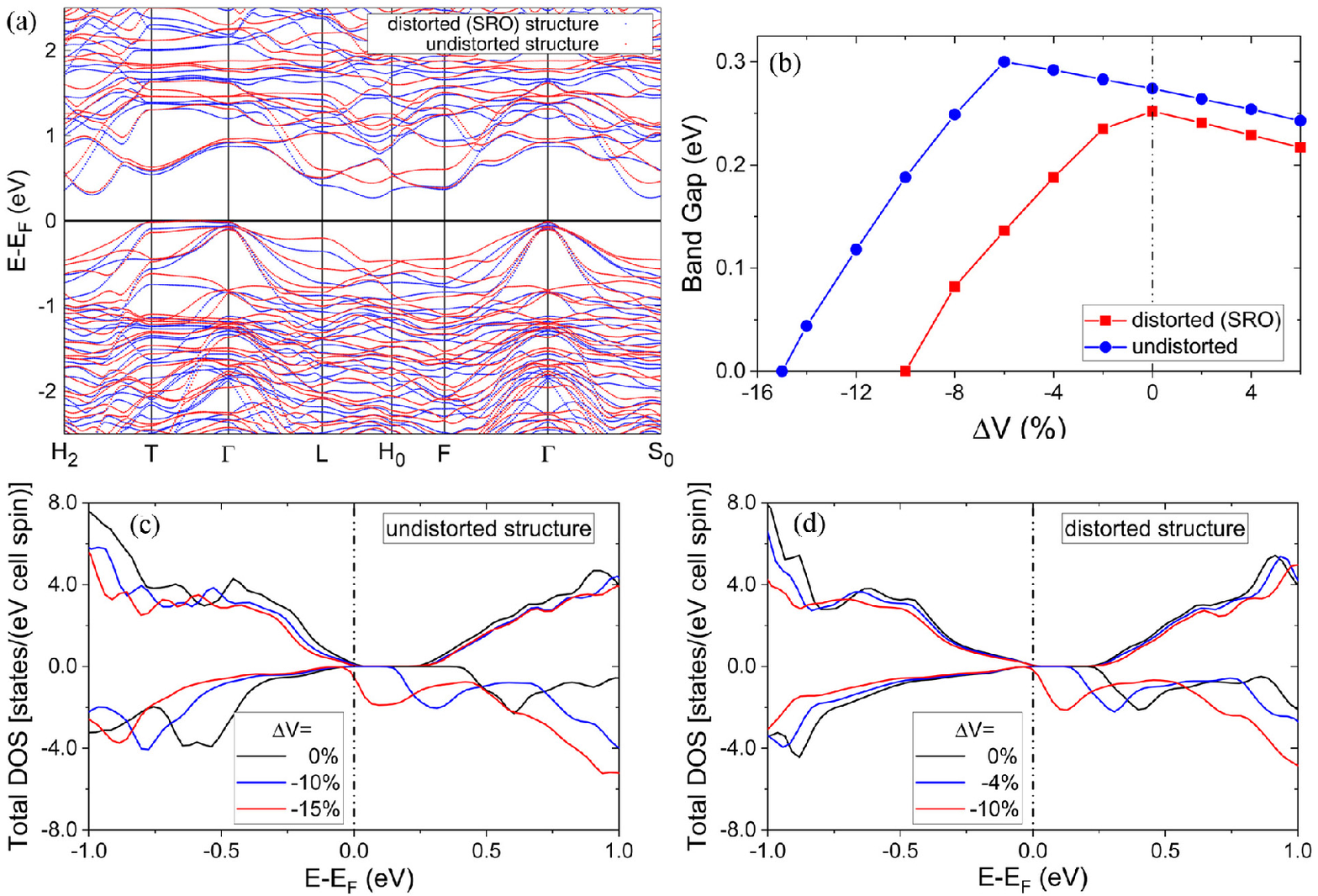}}
\caption{(Color online) (a) Comparison between band structures of distorted (SRO) and undistorted Cr$_2$Si$_2$Te$_6$ in the ambient pressure. (b) Electronic band gaps calculated as functions of volume change with pressure in distorted and undistorted Cr$_2$Si$_2$Te$_6$. The spin-orbit coupling is included. Spin-projected DOS calculated using different lattice parameters in (c) undistorted and (d) distorted structures. The minority spin states are more sensitive than the majority spin states to pressure, and the bandgap is determined by minority spin states.}
\end{figure*}


\begin{thebibliography}{99}
\bibitem{Jungwirth} T. Jungwirth, J. Sinova, J. Masek, J. Kucera, and A. H. MacDonald, Theory of ferromagnetic (III, Mn)V semiconductor. Rev. Mod. Phys. \textbf{78}, 809 (2006).
\bibitem{Lebegue} S. Leb\'{e}gue, T. Bj\"{o}rkman, M. Klintenberg, R. M. Nieminen, and O. Eriksson, Two-dimensional materials from data filtering and ab initio calculations. Phys. Rev. X \textbf{3}, 031002 (2013).
\bibitem{Li} X. Li and J. Yang, CrXTe$_3$ (X = Si, Ge) nanosheets: two dimensional intrinsic ferromagnetic semiconductors. J. Mater. Chem. C \textbf{2}, 7071 (2014).
\bibitem{Lin} M. Lin, H. Zhuang, J. Yan, T. Z. Ward, A. A. Puretzky, C. M. Rouleau, Z. Gai, L. Liang, V. Meunier, B. G. Sumpter, P. Ganesh, P. R. C. Kent, D. B. Geohegan, D. G. Mandrus, and K. Xiao, Ultrathin nanosheets of CrSiTe$_3$: a semiconducting two-dimensional ferromagnetic material. J. Mater. Chem. C \textbf{4}, 315 (2016).
\bibitem{Huang} B. Huang, G. Clark, E. Navarro-Moratalla, D. R. Klein, R. Cheng, K. L. Seyler, D. Zhong, E. Schmidgall, M. A. McGuire, D. H. Cobden, W. Yao, D. Xiao, P. Jarillo-Herrero, and X. D. Xu, Layer-dependent ferromagnetism in a van der Waals crystal down to the monolayer limit. Nature \textbf{546}, 270 (2017).
\bibitem{Gong} C. Gong, L. Li, Z. L. Li, H. W. Ji, A. Stern, Y. Xia, T. Cao, W. Bao, C. Z. Wang, Y. Wang, Z. Q. Qiu, R. J. Cava, S. G. Louie, J. Xia, and X. Zhang, Discovery of intrinsic ferromagnetism in two-dimensional van der Waals crystals. Nature \textbf{546}, 265 (2017).
\bibitem{RonA} A. Ron, S. Chaudhary, G. Zhang, H. Ning, E. Zoghlin, S. D. Wilson, R. D. Averitt, G. Refael, and D. Hsieh, Ultrafase enhancement of ferromagnetic spin exchange induced by ligand-to-metal charge transfer. Phys. Rev. Lett. \textbf{125}, 197203 (2020).
\bibitem{RalphD} D. C. Ralph and M. D. Stiles, Spin transfer torques. J. Magn. Magn. Mater. \textbf{320}, 1190 (2008).
\bibitem{BrataasA} A. Brataas, A. D. Kent, and H. Ohno, Current-induced torques in magnetic materials. Nat. Mater. \textbf{11}, 372 (2012).
\bibitem{WeiH} W. Han, Y. Otani, and S. Maekawa, Quantum materials for spin and charge conversion. npj Quantum Mater. \textbf{3}, 27 (2018).
\bibitem{ACSNANO} Q. H. Wang et al. The magnetic genome of two-dimensional van der Waals materials. ACS Nano \textbf{16} 6960 (2022).
\bibitem{Carteaux} V. Carteaux, F. Moussa, and M. Spiesser, 2D Ising-like ferromagnetic behaviour for the lameller Cr$_2$Si$_2$Te$_6$ compound: a neutron scattering investigation. EPL. \textbf{29}, 251 (1995).
\bibitem{Casto} L. D. Casto, A. J. Clune, M. O. Yokosuk, J. L. Musfeldt, T. J. Williams, H. L. Zhuang, M. W. Lin, K. Xiao, R. G. Hennig, B. C. Sales, J. Q. Yan, and D. Mandrus, Strong spin-lattice coupling in CrSiTe$_3$. APL Mater. \textbf{3}, 041515 (2015).
\bibitem{Siberchicot} B. Siberchicot, S. Jobic, V. Carteaux, P. Gressier, and G. Ouvrard, Band structure calculations of ferromagnetic chromium tellurides CrSiTe$_3$ and CrGeTe$_3$. Phys. J. Chem. \textbf{100}, 5863 (1996).
\bibitem{Zhang} X. Zhang, Y. Zhao, Q. Song, S. Jia, J. Shi, and W. Han, Magnetic anisotropy of the single-crystalline ferromagnetic insulator Cr$_2$Ge$_2$Te$_6$. Jpn. J. Appl. Phys. \textbf{55}, 033001 (2016).
\bibitem{Williams} T. J. Williams, A. A. Aczel, M. D. Lumsden, S. E. Nagler, and M. B. Stone, Magnetic correlations in the quasi-two-dimensional semiconducting ferromagnet CrSiTe$_3$. Phys. Rev. B \textbf{92}, 144404 (2015).
\bibitem{Milosavljevic} A. Milosavljevic, A. Solajic, B. Visic, M. Opacic, J. Pesic, Y. Liu, C. Petrovic, Z. V. Popovic, and N. Lazarevic, Vacancies and spin-phonon coupling in CrSi$_{0.8}$Ge$_{0.1}$Te$_3$. J. Raman Spectrosc. \textbf{51}, 2153-2160 (2020).
\bibitem{Cai} W. Cai, H. Sun, W. Xia, C. Wu, Y. Liu, H. Liu, Y. Gong, D. Yao, Y. Guo, and M. Wang, Pressure-induced superconductivity and structural transition in ferromagnetic CrSiTe$_3$. Phys. Rev. B \textbf{102}, 144525 (2020).
\bibitem{Song} C. Song, X. Liu, X. Wu, J. Wang, J. Pan, T. Zhao, C. Li, and J. Wang, Surface-vacancy-induced metallicity and layer-dependent magnetic anisotropy energy in Cr$_2$Ge$_2$Te$_6$. J. Appl. Phys. \textbf{126}, 105111 (2019).
\bibitem{Chen} X. Chen, J. Qi, and D. Shi, Strain-engineering of magnetic coupling in two-dimensional magnetic semiconductor CrSiTe$_3$: competition of direct exchange interaction and superexchange interaction. Phys. Lett. A \textbf{379}, 60 (2015).
\bibitem{Cheng} H. Cheng, J. Zhou, M. Yang, L. Shen, J. Linghu, Q. Wu, P. Qian, and Y. P. Feng, Robust two-dimensional bipolar magnetic semiconductors by defect engineering. J. Mater. Chem. C \textbf{6}, 8435 (2018).
\bibitem{LiX1} X. Li, X. Wu, and J. Yang, Room-temperature half-metallicity in La(Mn,Zn)AsO alloy via element substitutions. J. Am. Chem. Soc. \textbf{136}, 5664 (2014).
\bibitem{CarusoA} A. Caruso, K. I. Pokhodnya, W. W. Shum, W. Ching, B. Anderson, M. Brener, E. Vescovo, P. Rulis, A. Epstein, and J. S. Miller, Direct evidence of electron spin polarization from an organic-based magnet: [Fe$^{II}$(TCNE)(NCMe)$_2$][Fe$^{III}$Cl$_4$]. Phys. Rev. B \textbf{79}, 195202 (2009).
\bibitem{LiX2} X. Li and J. Yang, Bipolar magnetic materials for electrical manipulation of spin-polarization orientation. Phys. Chem. Chem. Phys. \textbf{15}, 15793 (2013).
\bibitem{ZhangJ} J. Zhang, X. Cai, W. Xia, A. Liang, J. Huang, C. Wang, L. Yang, H. Yuan, Y. Chen, S. Zhang, Y. Guo, Z. Liu, and G. Li, Unveiling electronic correlation and the ferromagnetic superexchange mechanism in the van der Waals crystal CrSiTe$_3$. Phys. Rev. Lett. \textbf{123}, 047203 (2019).
\bibitem{Ouvrard} G. Ouvrard, E. Sander, and R. Brec, Synthesis and crystal structure of a new layered phase: the chromium hexatellurosilicate Cr$_2$Si$_2$Te$_6$. J. Solid State Chem. \textbf{73}, 27 (1988).
\bibitem{Marsh} R. E. Marsh, The crystal strucutre of Cr$_2$Si$_2$Te$_6$. J. Solid State Chem. \textbf{77}, 190 (1988).
\bibitem{Carteaux2} V. Carteaux, D. Brunet, G. Ouvrard, and G. Andr\'{e}, Crystallographic, magnetic and electronic structures of a new layered ferromagnetic compound Cr$_2$Ge$_2$Te$_6$. J. Phys.: Condens. Matter \textbf{7}, 69 (1995).
\bibitem{Sivadas} N. Sivadas, M. W. Daniels, R. H. Swendsen, S. Okamoto, and D. Xiao, Magnetic ground state of semiconducting transition-metal trichalcogenide monolayers. Phys. Rev. B \textbf{91}, 235425 (2015).
\bibitem{Ofuchi} H. Ofuchi, N. Ozaki, N. Nishizawa, H. Kinjyo, S. Kuroda, and K. Takita, Fluorescence XAFS study on local structure around Cr atoms doped in ZnTe. AIP Conference Proceeding \textbf{882}, 517 (2007).
\bibitem{AFM} Y. Liu, M.-G. Han, Y. Lee, M. O. Ogunbunmi, Q. Du, C. Nelson, Z. Hu, E. Stavitski, D. Graf, K. Attenkofer, S. Bobev, L. Ke, Y. Zhu, and C. Petrovic, Polaronic conductivity in Cr$_2$Ge$_2$Te$_6$ single crystals. Adv. Funct. Mater. \textbf{2105111}, 2105111 (2022).
\bibitem{YLPRB} Y. Liu, M. Abeykoon, E. Stavitski, K. Attenkofer, and C. Petrovic, Magnetic anisotropy and entropy change in trigonal Cr$_5$Te$_8$. Phys. Rev. B \textbf{100}, 245114 (2019).
\bibitem{Prins} R. Prins and D. C.Koningsberger (eds.), X-ray Absorption: Principles, Applications, Techniques of EXAFS, SEXAFS, XANES (Wiley, New York, 1988).
\bibitem{Carteaux3} V. Carteaux, G. Ouvrard, J. C. Grenier, and Y. Laligant, Magnetic structure of the new layered ferromagnetic chromium hexatellurosilicate Cr$_2$Si$_2$Te$_6$. J. Magn. Magn. Mater. \textbf{94}, 127-133 (1991).
\bibitem{Hao} Z. Hao, H. Li, S. Zhang, X. Li, G. Lin, X. Luo, Y. Sun, Z. Liu, and Y. Wang, Atomic scale electronic structure of the ferromagnetic semiconductor Cr$_2$Ge$_2$Te$_6$. Science Bulletin \textbf{63}, 825 (2018).
\bibitem{CaoY} Y. Cao, V. Fatemi, S. Fang, K. Watanabe, T. Taniguchi, E. Kaxiras, and P. Jarillo-Herrero, Unconventional superconductivity in magic-angle graphene superlattices. Nature \textbf{556}, 43-50 (2018).
\bibitem{Bomerich} T. B\"{o}merich, L. Heinen, and A. Rosch, Skymion and tetarton lattices in twisted bilayer graphene. Phys. Rev. B \textbf{102}, 100408 (2020).
\bibitem{YuLiu} Y. Liu and C. Petrovic, Anisotropic magnetic entropy change in Cr$_2$X$_2$Te$_6$ (X = Si and Ge). Phys. Rev. Mater. \textbf{3}, 014001 (2019).
\bibitem{XieQ} Q. Xie, Y. Liu, M. Wu, H. Lu, W. Wang, L. He, and X. Wu, Two state magnetization in van der Waals layered CrXTe$_3$ (X = Si, Ge) single crystals. Mater. Lett. \textbf{246}, 60 (2019).
\bibitem{Gopal} E. S. R. Gopal, Specific Heats at Low Temperatures (Plenum Press, New York, 1966).
\bibitem{Hentrich} R. Hentrich, A. U. B. Wolter, X. Zotos, W. Brenig, D. Nowak, A. Isaeva, T. Doert, A. Banerjee, P. Lampen-Kelley, D. G. Mandrus, S. E. Nagler, J. Sears, Y.-J. Kim, B. B\"{u}chner, and C. Hess, Unusual phonon heat transport in $\alpha$-RuCl$_3$: strong spin-phonon scattering and field-induced spin gap. Phys. Rev. Lett. \textbf{120}, 117204 (2018).
\bibitem{XuC1} C. Xu, J. Feng, H. Xiang, and L. Bellaiche, Interplay between Kitaev interaction and single ion anisotropy in ferromagnetic CrI$_3$ and CrGeTe$_3$ monolayers. npj Comput. Mater. \textbf{4}, 57 (2018).
\bibitem{XuC2} C. Xu, J. Feng, M. Kawamura, Y. Yamaji, Y. Nahas, S. Prokhorenko, Y. Qi, H. Xiang, and L. Bellaiche, Possible Kitaev quantum spin liquid state in 2D materials with $S$ = 3/2. Phys. Rev. Lett. \textbf{124}, 087205 (2020).
\bibitem{Austin} I. G. Austin, and N. F. Mott, Polarons in crystalline and non-crystalline materials. Adv. Phys. \textbf{50}, 757 (2001).
\bibitem{Zvyagin} I. P. Zvyagin, On the theory of hopping transport in disordered semiconductors. Phys. Status Solidi B \textbf{58}, 443 (1973).
\bibitem{Lee} P. A. Lee and T. V. Ramakrishnan, Disordered electronic systems. Rev. Mod. Phys. \textbf{57}, 287 (1985).
\bibitem{SinghS} S. Singh, S. Pal, and C. Biswas, Disorder induced resistivity anomaly in Ni$_2$Mn$_{1+x}$Sn$_{1-x}$. J. Alloy. Comp. \textbf{616}, 110 (2014).
\bibitem{Yu} Z. Yu, W. Xia, K. Xu, M. Xu, H. Wang, X. Wang, N. Yu, Z. Zou, J. Zhao, L. Wang, X. Miao, and Y. Guo, Pressure-induced structural phase transition and a special amorphization phase of two-dimensional ferromagnetic semiconductors Cr$_2$Ge$_2$Te$_6$. J. Phys. Chem. C \textbf{123}, 13885 (2019).
\bibitem{mott} N. F. Mott, Metal-Insulator Transitions (Taylor and Francis, London, 1990).
\bibitem{kurosaki} Y. Kurosaki, Y. Shimizu, K. Miyagame, K. Kanoda and G. Saito, Mott transition from a spin liquid to a Fermi liquid in the spin-frustrated organic conductor $\kappa$-(ET)$_2$Cu$_2$(CN)$_3$. Phys. Rev. Lett., \textbf{95}, 177001 (2005).
\bibitem{xu} K. Xu,  Z. Yu, W. Xia, M. Xu, X. Mai, L. Wang, Y. Guo, X. Miao, and M. Xu, Unique 2D-3D structure transformation in trichalcogenide CrSiTe$_3$ under high pressure. J. Phys. Chem. C \textbf{124}, 15600 (2020).
\bibitem{milosav}A. Milosavljevic, A. Solajic, J. Pesic, Yu Liu, C. Petrovic, N. Lazarevic, and Z. V. Popovic, Evidence of spin-phonon coupling in CrSiTe$_3$. Phys. Rev. B. \textbf{98}, 104306 (2018).
\bibitem{lee2020prb} Y. Lee, T. Kotani, and L. Ke, Role of nonlocality in exchange correlation for magnetic two-dimensional van der Waals materials. Phys. Rev. B \textbf{101}, 241409 (2020).
\bibitem{Murnaghan244} F. D. Murnaghan, The compressibility of media under extreme pressures. Proceedings of the National Academy of Sciences \textbf{30}, 244 (1944).
\bibitem{Menichetti2019_2dmat} G. Menichetti, M. Calandra, and M. Polini, Electronic structure and magnetic properties of few-layer Cr$_2$Ge$_2$Te$_6$: the key role of nonlocal electron-electron interaction effects. 2D Materials \textbf{6}, 045042 (2019).
\bibitem{mcguire2015cm} M. A. McGuire, H. Dixit, V. R. Cooper, and B. C. Sales, Coupling of crystal strucuture and magnetism in the layered. ferromagnetic insulator CrI$_3$. Chem. Mater. \textbf{27}, 612 (2015).
\bibitem{ZouX} X. Zou and B. I. Yakobson, An open canvas-2D materials with defects, disorder, and functionality. Acc. Chem. Res. \textbf{48}, 73 (2015).
\bibitem{Kieffer} J. Kieffer, V. Valls, N. Blanc, and C. Hennig, New tools for calibrating diffraction setups. Journal of Synchrotron Radiation \textbf{27}, 558 (2020).
\bibitem{Yang} X. Yang, P.Juhas, C. L. Farrow, and S. J. L. Billinge, xPDFsuite: an end-to-end software solution for high throughput pair distribution function transformation, visualization and analysis. (2015) arXiv:1402.3163. http://doi.org/10.48550/arXiv.1402.3163 (23 Feb. 2015).
\bibitem{Toby} B. H. Toby and R. B. Von Dreele, ``GSAS-II: the genesis of a modern open-source all purpose crystallography soft- ware package," Journal of Applied Crystallography \textbf{46}, 544 (2013).
\bibitem{Farrow} C. L. Farrow, P. Juhas, J. W. Liu, D. Bryndin, E. S. Bozin, J. Bloch, Th. Proffen, and S. J. L. Billinge, PDFfit2 and PDFgui: computer programs for studying nanostructure in crystals, J. Phys.: Condens. Matter \textbf{19}, 335219 (2007).
\bibitem{mao} H. K. Mao, J. Xu, and P. M. Bell, Calibration of the ruby pressure gauge to 800 kbar under quasi-hydrostatic conditions. J. Geophys. Res. \textbf{91}, 4673-4676 (1986).
\bibitem{wien2k} P. Blaha, K. Schwarz, G. K. H. Madsen, D. Kvasnicka, J. Luitz, R. Laskowski, F. Tran, and L. D. Marks, WIEN2k: An Augmented Plane Wave plus Local Orbitals Program for Calculating Crystal Properties (Vienna University of Technology, Austria, 2018).
\bibitem{perdew1996prl} J. P. Perdew, K. Burke, and M. Ernzerhof, Generalized Gradient Approximation Made Simple, Phys. Rev. Lett. \textbf{77}, 3865 (1996).
\bibitem{madsen2006cpc} G. K. Madsen and D. J. Singh, BoltzTraP. A code for calculating band-structure dependent quantities, Computer Physics Communications \textbf{175}, 67 (2006).
\end{thebibliography}
\end{document}